\documentclass{WileyMSP-template}
\usepackage{amsmath}
\usepackage{hyperref}
\usepackage{hyperref}  
\usepackage[numbers,sort&compress]{natbib}

\begin{document}

\pagestyle{fancy}

\title{Non-Markovian dynamics with  $\Lambda$-type atomic systems in a single-end photonic waveguide}
	
\maketitle

	
\author{Jun-Cong Zheng}
\author{Xiao-Wei Zheng}
\author{Xin-Lei Hei}
\author{Yi-Fan Qiao}
\author{Xiao-Yu Yao}
\author{Xue-Feng Pan}
\author{Yu-Meng Ren}
\author{Xiao-Wen Huo}
\author{Peng-Bo Li}

	

\begin{affiliations}

Ministry of Education Key Laboratory for Nonequilibrium Synthesis and
Modulation of Condensed Matter
Shaanxi Province Key Laboratory of Quantum Information and Quantum
Optoelectronic Devices
School of Physics
Xi’an Jiaotong University
Xi’an 710049, China\\

\end{affiliations}


\begin{abstract}
		
In this work, we investigate the non-Markovian dynamical evolution of a $\Lambda$-type atom interacting with a semi-infinite one-dimensional photonic waveguide via two atomic transitions. The waveguide terminates at a perfect mirror, which reflects the light and introduces boundary effects. We derive exact analytical expressions and show that, under suitable conditions, the instantaneous and retarded decay rates reach equilibrium, leading to the formation of an atom-photon bound state that suppresses dissipation. Consequently, the atom retains a long-lived population in the asymptotic time limit. Furthermore, we analyze the output field intensity and demonstrate that blocking one of the coupling channels forces the atomic system to emit photons of a single frequency. Finally, we extend the model to a two-atom system and examine the disentanglement dynamics of the two spatially separated atoms. These findings elucidate the dynamic process of spontaneous emission involving multi-frequency photons from multi-level atoms and provide insights into the complex interference between different decay pathways.
		
\end{abstract}
	
	
\section{Introduction}\label{I}
	
~~~~~The interaction between light and matter serves as a fundamental model for exploring quantum dynamics between atomic systems and the electromagnetic field, playing a pivotal role in advancing quantum science over the past decades \cite{PhysRevB.5.648,weiner2008light,frisk2019ultrastrong,RevModPhys.91.025005,rivera2020light}. By introducing geometric constraints \cite{PhysRev.69.37,miller2005trapped,RevModPhys.73.565,PhysRevLett.64.2418}, researchers can investigate specific effects in single-mode cases, leveraging platforms such as cavity quantum electrodynamics (QED) systems \cite{mabuchi2002cavity,PhysRevA.69.062320,PhysRevA.97.043820,mirhosseini2019cavity,haroche2020cavity,owens2022chiral} and circuit-QED systems \cite{PhysRevA.75.032329,niemczyk2010circuit,RevModPhys.93.025005,blais2020quantum,clerk2020hybrid}.
In these systems, spontaneous emission causes an excited atom to decay to its ground state by emitting single-mode light into the environment. Moreover, one-dimensional (1D) waveguide QED systems have emerged as promising platforms for quantum information processing and quantum computation \cite{PhysRevLett.111.090502,RevModPhys.95.015002,PhysRevResearch.5.023031,PRXQuantum.4.030326,PhysRevX.10.031011,zheng2023few,PhysRevA.109.063709}, typically supporting a continuum of bosonic modes.	A variety of artificial systems have been proposed and experimentally realized to implement light-matter interactions in 1D waveguides, including photonic crystal waveguides \cite{goban2014atom,PhysRevLett.115.063601}, superconducting qubits coupled to microwave transmission lines \cite{PhysRevA.77.013831,PhysRevA.89.053802}, and surface acoustic waves \cite{gustafsson2014propagating}. These experimental advancements in atom-waveguide systems offer unprecedented opportunities to investigate  interference across multiple coupling pathways.

~~~~~Interference is a common phenomenon in giant atom systems, which feature nonlocal interactions between atoms and waveguide fields \cite{gustafsson2014propagating,satzinger2018quantum,PhysRevLett.124.240402,PhysRevA.103.023710}. To date, various intriguing phenomena have been observed in giant-atom structures, including decoherence-free interatomic interactions \cite{kannan2020waveguide,PhysRevLett.120.140404,PhysRevResearch.2.043184,PhysRevResearch.2.043070,PhysRevA.105.023712}, unconventional bound states \cite{PhysRevLett.126.043602,PhysRevResearch.2.043014,PhysRevA.102.033706}, and frequency-dependent relaxation rates and Lamb shifts \cite{PhysRevA.90.013837,PhysRevA.104.033710}.
Interestingly, interference effects can also occur in point-like atom systems when the waveguide is terminated with a mirror. In a setup consisting of a single atom in front of a mirror—extensively studied both experimentally \cite{hoi2015probing,eschner2001light,PhysRevLett.91.213602,PhysRevLett.98.183003} and theoretically \cite{PhysRevA.91.053845,PhysRevLett.116.093601,PhysRevA.92.053834,PhysRevA.41.1587,PhysRevA.66.023816,PhysRevA.66.063801,PhysRevA.79.063847,koshino2012control,PhysRevA.85.013823,PhysRevA.87.013820,PhysRevA.105.053706,PhysRevA.109.023712}—decay can be suppressed through interference between the relaxation pathways of the atom and its mirror image.
Moreover, such systems provide an intriguing platform to study non-Markovian retardation effects. In these cases, photons act as transmission resources, and the propagation time between coupling points becomes comparable to or even exceeds the atom's lifetime \cite{andersson2019non}. However, while these studies primarily focus on the interference of single-mode light, the interference effects involving multimode light remain largely unexplored.

~~~~~In this work, we consider a $\Lambda$-type atom coupled to a 1D semi-infinite waveguide through two atomic transitions, with potentially two light modes traveling in the system. A mirror is placed on one side of the waveguide, which may reflect photons moving towards it and thereby induce time-delayed quantum dynamics. The time taken by the emitted photon to complete a round trip between the atom and the mirror serves as the memory time of the system.
We derive an exact delay differential equation for the atomic excitation amplitude. One distinguishing feature compared to a two-level atomic system is that the suppression of spontaneous emission depends on the balance between the instantaneous and retarded decay rates of the two modes. By appropriately adjusting the distance between the atom and the mirror, one of these transmission channels can be blocked, allowing the release of photons with a specific single frequency. This prediction is tested by analyzing the output field intensity. Additionally, we demonstrate that introducing an atomic frequency shift can release the trapped excitation, initially in a bound atom-photon stationary state.
Finally, we extend the model to the two-atom case to explore the disentanglement dynamics of two separate atoms.

~~~~~This work is organized as follows: In Section \ref{II}, we introduce the model of the $\Lambda$-type atom in a semi-infinite waveguide system and describe the methods used to address the Non-Markovian dynamics.
In Section \ref{III}, we derive the delay differential equation for the two modes governing the time evolution of the atomic excitation.
Section \ref{IV} focuses on the derivation of the bound state for the total system.
In Section \ref{V}, we examine the output field intensity to further validate the spontaneous emission dynamics of the $\Lambda$-type atom.
In Section \ref{VI}, we extend the discussion to a general model involving a non-ideal mirror.
In Section \ref{VII}, we generalize the system to include two $\Lambda$-type atoms coupled to a one-dimensional semi-infinite waveguide, deriving a set of delay differential equations for the probability amplitudes.
Finally, Section \ref{VIII} concludes with a summary of our findings.

\section{The Model and Methods}\label{II}

\begin{figure}[htbp]
	\centering
	\resizebox{0.4\columnwidth}{!}{
			\includegraphics{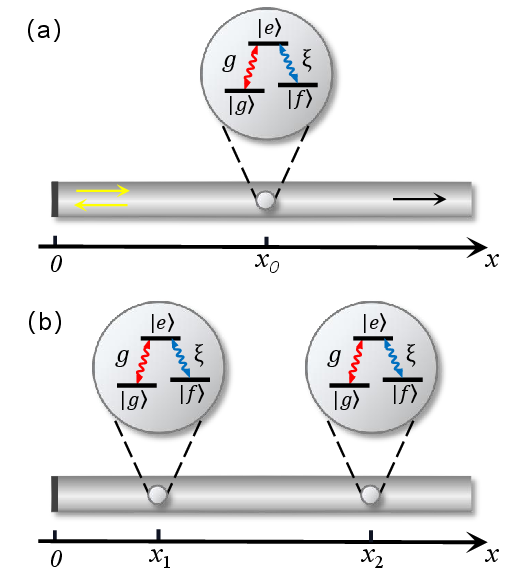}	}
		\renewcommand\figurename{\textbf{FIG.}}
		\caption[1]{Schematic of the setup. (a) At $x = x_0$, a $\Lambda$-type atom is coupled to a semi-infinite waveguide, where the atomic transition $\vert g \rangle \leftrightarrow \vert e \rangle$ is coupled to the waveguide with a coupling strength $g$, and the atomic transition $\vert f \rangle \leftrightarrow \vert e \rangle$ is coupled to the waveguide with a coupling strength $\xi$. The transition frequencies between the energy states are $\omega_0$ and $\omega_0-\omega_f$, corresponding to wave vector $k$ and wave vector $q$ in the waveguide, respectively. (b) Two identical $\Lambda$-type atoms as shown in (a) are coupled to the semi-infinite waveguide at $x = x_1$ and $x = x_2$, respectively.}
		\label{fig.1}
\end{figure}

~~~~~As shown in Figure \ref{fig.1}(a), our system consists of a $\Lambda$-type atom coupled to a 1D semi-infinite waveguide at $x = x_0$, where the two atomic transitions are coupled to the waveguide with coupling strengths $g$ and $\xi$, respectively. The semi-infinite waveguide can be realized using photonic-crystal waveguides \cite{wang2018classical,wang2018role,lodahl2004controlling} or microwave transmission lines \cite{gu2017microwave,PhysRevA.84.063834,PhysRevA.101.053861,PhysRevLett.112.170501,PhysRevA.86.032106}, and it is terminated by a perfect mirror at $x = 0$ with reflectivity $R = 1$.
Considering the spontaneous emission of the atom, the photons radiated by the atom and reflected by the mirror can coherently interact with subsequent emitted photons. This phenomenon is analogous to the coherent interaction between the two coupled arms of a giant atom in an infinite waveguide \cite{gustafsson2014propagating,satzinger2018quantum,PhysRevLett.124.240402,PhysRevA.103.023710}.
The three-level atom has $\omega_0$ and $\omega_f$ as the energies of the states $\vert e \rangle$ and $\vert f \rangle$ relative to the ground state $\vert g \rangle$, respectively. In this work, we denote the annihilation (creation) operator $a_{\mu} (a_{\mu}^{\dagger})$ ($\mu = k, q$) to describe the photon in the waveguide with wave vector ${\mu}$ and frequency $\omega_{\mu}$, which obey the bosonic commutation rule $[a_{\mu}, a_{{\mu}^{'}}^{\dagger}] = \delta({\mu} - {\mu}^{'})$.
Under the Weisskopf-Wigner approximation, the dispersion relation for photons with wave vectors $k$ and $q$ in the waveguide is approximately linear around the transition frequencies, expressed as $\omega_{k} \simeq \omega_0 + (\kappa - \kappa_0)v_k$ and $\omega_{q} \simeq \omega_0 - \omega_f + (q - q_0)v_q$ \cite{PhysRevA.79.023837,PhysRevLett.95.213001}, where $v_k$ and $v_q$ are the photon group velocities, and $k_0$ and $q_0$ satisfy $\omega_{k_0} = \omega_0$ and $\omega_{q_0} = \omega_0 - \omega_f$.
In general, two orthogonal standing modes with spatial profiles proportional to $\sin(\mu x)$ and $\cos(\mu x)$ are possible. However, for the semi-infinite waveguide, only the sine-like modes are considered. Consequently, the atom is dipole-coupled to mode $k$ and mode $q$ with coupling strengths $g \propto \sin(k x_0)$ and $\xi \propto \sin(q x_0)$, respectively.
Under the rotating-wave approximation, the Hamiltonian of the system is expressed as (with $\hbar = 1$):
	
\begin{eqnarray}\label{eq1} 
	H &=&\omega_0\vert e\rangle\langle e\vert +\omega_f\vert f\rangle\langle f\vert +\sum_{\mu=k,q}\int_{0}^{\mu_c} d\mu ~\omega_\mu a_\mu^{\dagger}a_\mu\nonumber \\&&+(\int_{0}^{\kappa_c} dk ~g\vert e\rangle\langle g\vert a_k+\int_{0}^{q_c} dq ~\xi\vert e\rangle\langle f\vert a_q+H.c.),
\end{eqnarray}
where $\mu_c$ represents the cutoff wave vector, which depends on the specific characteristics of the waveguide. 
	
~~~~~We consider the $\Lambda$-type atom to be initially in the excited state $\vert e\rangle$, while the field in the waveguide remains in the vacuum state $\vert 0\rangle$, represented collectively as $\vert e,0\rangle$. During the process of atomic spontaneous emission, the atom emits a photon with wave vector $k$ or $q$ into the waveguide, accompanied by a transition of the atomic system to the ground state $\vert g\rangle$ or the metastable state $\vert f\rangle$, respectively.	The total number of excitations is conserved, as indicated by $[H,\vert e\rangle\langle e\vert+\sum_{\mu=k,q}\int d\mu~a_{\mu}^{\dagger}a_{\mu}]=0$. Therefore, the state of the total system in the single-excitation subspace can be expressed as
\begin{eqnarray}\label{eq2} 
		\vert \psi(t) \rangle &=&\int dk~ c_k(t)     
		a_k^{\dagger}\vert g,0\rangle+\int dq~ c_q(t)     
		a_q^{\dagger}\vert f,0\rangle+c_e(t)\vert e,0\rangle,\nonumber \\&&
\end{eqnarray}
where $c_e(t)$ is the atomic excitation probability amplitude, and $c_{\mu}(t)$ is the field amplitude in the $\mu$-space. By solving the time-dependent Schrödinger equation $i\partial_t \vert \psi \rangle = H \vert \psi \rangle$, one obtains
	
\begin{equation}\label{eq3}
	\begin{aligned}
			&\dot{c}_{e}(t)=-i\omega_0 c_{e}(t)-i\int_{0}^{\kappa_c} dk~ gc_k(t)-i\int_{0}^{q_c} dq~ \xi c_q(t),\\
			&\dot{c}_{k}(t)=-i\omega_k c_{k}(t)-ig^{*} c_{e}(t),\\
			&\dot{c}_{q}(t)=-i(\omega_q+\omega_f) c_{q}(t)-i\xi^{*} c_{e}(t).\\
	\end{aligned}
\end{equation}

~~~~~To simplify our calculations, we approximate the integral bounds as $\int_{0}^{\mu_c} d\mu \rightarrow \int_{-\infty}^{+\infty} d\mu$ \cite{RevModPhys.89.021001,shen2005coherent,PhysRevA.101.032335,liao2016photon}, and set the coupling strengths as $g = \sqrt{\Gamma_g v_k/\pi} ~\mathrm{sin}(\kappa x_0)$ and $\xi = \sqrt{\Gamma_{\xi} v_q/\pi} ~\mathrm{sin}(q x_0)$ \cite{PhysRevA.90.012113}, where $\Gamma_g$ and $\Gamma_{\xi}$ represent the spontaneous emission rates of the atom in the absence of a mirror.

\section{Non-Markovian Dynamics of  a $\Lambda$-type Atom}\label{III}
	
~~~~~For the case where the waveguide field is initialized in the vacuum state (i.e., $c_e(0) = 1$, $c_k(0) = c_q(0) = 0$), applying the transformations $c_{e}(t) \rightarrow c_{e}(t) e^{-i\omega_0 t}$, $c_{k}(t) \rightarrow c_{k}(t) e^{-i\omega_0 t}$, and $c_{q}(t) \rightarrow c_{q}(t) e^{-i\omega_0 t}$ to Eq. (\ref{eq3}), the formal solutions for $c_k(t)$ and $c_q(t)$ can be written as
	
\begin{equation}\label{eq4}
	\begin{aligned}
			&c_{k}(t)=-i\int_0^{t} dt^{'}e^{-iv_k(k-k_0)(t-t^{'})}\times\sqrt{\Gamma_g v_k/\pi} ~\mathrm{sin}(\kappa x_0)c_{e}(t^{'}),\\
			&c_{q}(t)=-i\int_0^{t} dt^{'}e^{-iv_q(q-q_0)(t-t^{'})}\times\sqrt{\Gamma_{\xi} v_q/\pi} ~\mathrm{sin}(q x_0)c_{e}(t^{'}).
	\end{aligned}
\end{equation}
	
~~~~~Substituting Eq.(\ref{eq4}) into the dynamic equation of $c_e(t)$ in Eq.(\ref{eq3}), there is
	
\begin{equation}\label{eq5}
	\begin{aligned}
			&\dot{c}_{e}(t)=-\frac{\Gamma_g v_k}{\pi}\int_0^t dt^{'}c_{e}(t^{'})e^{ik_0 v_k(t-t^{'})}\int_{-\infty}^{+\infty} dk~ \mathrm{sin}^2(\kappa x_0)\times e^{-ik v_k(t-t^{'})}\\
			&~~~~~~~~~~~-\frac{\Gamma_{\xi} v_q}{\pi}\int_0^t dt^{'}c_{e}(t^{'})e^{iq_0 v_q(t-t^{'})}\int_{-\infty}^{+\infty} dq~ \mathrm{sin}^2(q x_0)\times e^{-iq v_q(t-t^{'})}.\\
	\end{aligned}
\end{equation}
After integrating over the wave vector $\mu$ and time $t'$ in Eq. (\ref{eq5}), one can obtain a delay differential equation for $c_e(t)$ (see \hyperref[Appendix]{Appendix} for more details):
\begin{equation}\label{eq6}
	\begin{aligned}
			&\dot{c}_{e}(t)=-\frac{\Gamma_g}{2}c_{e}(t)+\frac{\Gamma_g}{2}e^{i\phi_k}c_{e}(t-\tau_k)\Theta(t-\tau_k)\\
			&~~~~~~~~~-\frac{\Gamma_{\xi}}{2}c_{e}(t)+\frac{\Gamma_{\xi}}{2}e^{i\phi_q}c_{e}(t-\tau_q)\Theta(t-\tau_q),
	\end{aligned}
\end{equation}
where $\Theta(t)$ denotes the Heaviside step function, which describes the delayed feedback from the coupling point and the reflection from the mirror in the waveguide. The phase is given by $\phi_{\mu} = 2\mu_0 x_0$. The time delay $\tau_{\mu} = 2x_0/v_{\mu}$ represents the time taken by a photon with wave vector $\mu$ to perform a round trip between the atom and the mirror. The first and third terms on the right-hand side of Eq. (\ref{eq6}) describe the standard damping at rates $\Gamma_g$ and $\Gamma_{\xi}$ through the atomic transitions $\vert g \rangle \leftrightarrow \vert e \rangle$ and $\vert f \rangle \leftrightarrow \vert e \rangle$, respectively. The second and fourth terms, on the other hand, indicate that atomic reabsorption of the emitted photon can occur at times $t \geq \tau_{\mu}$.
Clearly, the non-Markovian dynamics of the excited $\Lambda$-type atom in our model are influenced by two atomic transition channels coupled to the waveguide, including factors such as the coupling strengths, phases, and time delays. We will analyze these contributing factors in the following discussion.

~~~~~Before proceeding, we briefly revisit the typical dynamics of a time-independent $\Lambda$-type atom described by Eq. (\ref{eq6}). It has been shown that the spontaneous emission of such an atom can be suppressed if $\phi_{\mu} = 2n\pi$ (the atom-waveguide coupling path interferes destructively with the part reflected by the mirror), and if the propagation time $\tau_{\mu}$ is negligible compared with the lifetime of the atom (i.e., the system is in the Markovian regime). In this case, the atom is effectively decoupled from the waveguide and becomes decoherence-free \cite{PhysRevLett.120.140404,kannan2020waveguide}. On the other hand, the atom can also exhibit superradiance behavior if the two-photon scattering paths interfere constructively.
We first assume that the two lower states $\vert g \rangle$ and $\vert f \rangle$ of the three-level atom are quasi-degenerate \cite{PhysRevResearch.3.043226}, such that the group velocities corresponding to the two wave vectors are identical (i.e., $v_k = v_q$), which results in the time delay $\tau_k = \tau_q = \tau$. By proceeding similarly to Ref. \cite{PhysRevA.59.2524}, Eq. (\ref{eq6}) can be solved iteratively by partitioning the time axis into intervals of length $\tau$, and its explicit solution is given by
\begin{eqnarray}\label{eq7}
	c_e(t) &=& \sum_{n=0}^{\infty} \left( \dfrac{\Gamma_g}{2} e^{i\phi_k} + \dfrac{\Gamma_{\xi}}{2} e^{i\phi_q} \right)^n \frac{(t - n\tau)^n}{n!}\nonumber \\&&
	\times e^{ - \scriptstyle \frac{\Gamma_g + \Gamma_{\xi}}{2}(t - n\tau) } \Theta(t - n\tau).
\end{eqnarray}
	
\begin{figure}[htbp]
		\centering
		\resizebox{0.4\columnwidth}{!}{
			\includegraphics{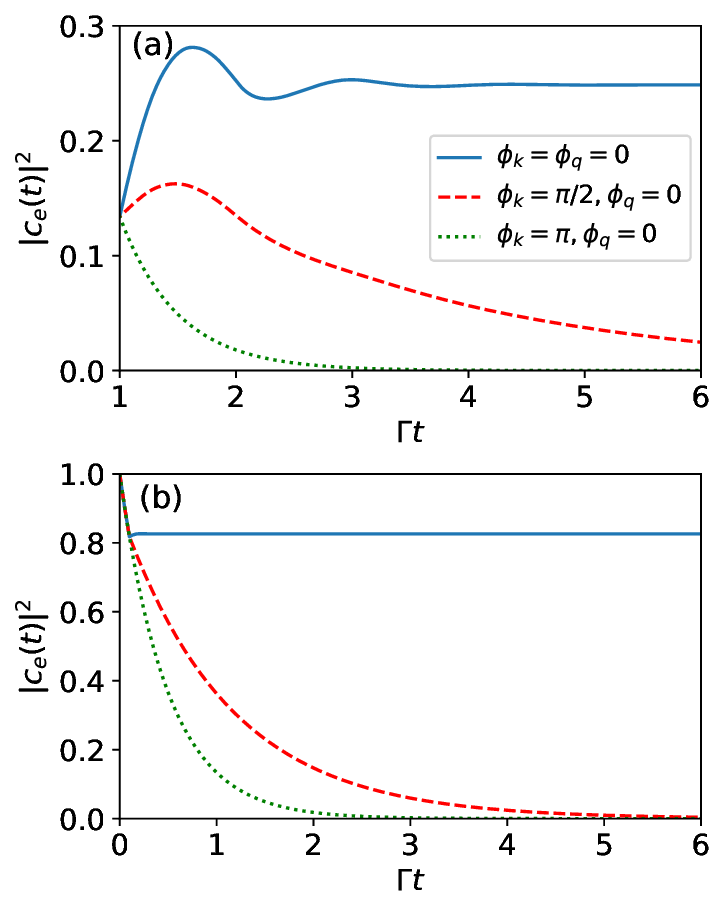}
		}
		\renewcommand\figurename{\textbf{FIG.}}
		\caption[2]{The probability amplitudes given by Eq. (\ref{eq6}) as functions of time $t$ (in units of $\Gamma^{-1}$) are shown for the cases $\Gamma\tau_k = \Gamma\tau_q = 1$ (a) and $\Gamma\tau_k = \Gamma\tau_q = 0.1$ (b). The different curves represent the following phase conditions: $\phi_k = \phi_q = 2n\pi$ (solid blue line), $\phi_k = \pi/2 + 2n\pi, \phi_q = 2n\pi$ (red dashed line), and $\phi_k = (2n+1)\pi, \phi_q = 2n\pi$ (green dotted line). In panel (a), only the range $t \geq \tau_{\mu}$ is shown; for earlier times, the behavior does not depend on $\phi_{\mu}$. The other parameter chosen is $\Gamma_g = \Gamma_{\xi} = \Gamma$. }
		\label{fig.2}
\end{figure}
	
~~~~~In Fig. \ref{fig.2}, we plot the time evolution of the atomic excitation probability $|c_e(t)|^2$ for different values of $\phi_{\mu}$, for $\Gamma\tau_k = \Gamma\tau_q = 1$ (a) and $\Gamma\tau_k = \Gamma\tau_q = 0.1$ (b). Before $t = \tau_{\mu}$, the excited atom interacts with the waveguide through two coupling channels, accompanied by the spontaneous emission of photons with wave vectors $k$ and $q$ into the waveguide simultaneously. As soon as $t \geq \tau_{\mu}$, the presence of the mirror starts affecting the atom, causing the dynamics to become strongly dependent on $\phi_{\mu}$ and $\tau_{\mu}$. Only for $\phi_k = \phi_q = 2n\pi$ does the atom reach a state where spontaneous emission is suppressed (i.e., in the long time limit, as $t \to \infty$, the population of the atom does not vanish.), and the final population of the atom is related to the time delay $\tau_{\mu}$. To demonstrate this, we take the Laplace transform (LT) of Eq. (\ref{eq6}) and solve the resulting algebraic equation. This yields
\begin{equation}\label{eq8}
	\begin{aligned}
			&\widetilde{c}_{e}(s)=\dfrac{c_{e}(0)}{s+\dfrac{\Gamma_g}{2}(1-e^{i\phi_k-s\tau_k})+\dfrac{\Gamma_{\xi}}{2}(1-e^{i\phi_q-s\tau_q})},
	\end{aligned}
\end{equation}
where $\widetilde{c}_{e}(s)$ is the LT of $c_e(t)$. Using the final value theorem \cite{gluskin2003let}
\begin{equation}\label{eq9}
		c_e(t\rightarrow\infty)=\lim\limits_{s \to 0}
		[s\widetilde{c}_{e}(s)].
\end{equation}
Substituting Eq.(\ref{eq8}) into Eq.(\ref{eq9}), we obtain the relation
\begin{equation}\label{eq10}
		\dfrac{\Gamma_g}{2}(1-e^{i\phi_k})+ \dfrac{\Gamma_{\xi}}{2}(1-e^{i\phi_q})=0.
\end{equation}
The solution determined by this condition is given by $\phi_k = \phi_q = 2n\pi$. The steady-state of the $\Lambda$-type atom is then obtained as
\begin{equation}\label{eq11}
		|c_e(t \to \infty)|^2=(1+\dfrac{\Gamma_g\tau_k+\Gamma_{\xi}\tau_q}{2})^{-2},  ~~~\phi_k=\phi_q=2n\pi.
\end{equation}
\begin{figure}[htbp]
		\centering
		\resizebox{0.4\columnwidth}{!}{
			\includegraphics{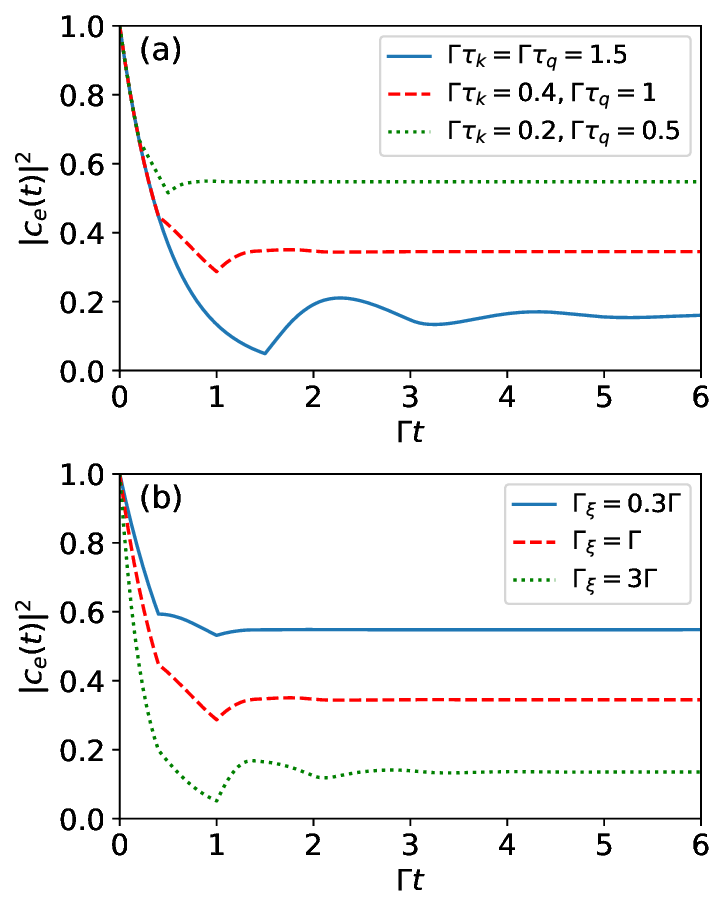}
		}
		\renewcommand\figurename{\textbf{FIG.}}
		\caption[3]{Probability amplitudes $|c_e(t)|^2$ as functions of time $t$ (in units of $\Gamma^{-1}$) are shown in the following cases: In panel (a), for $\Gamma_g = \Gamma_{\xi} = \Gamma$, the cases are $\Gamma\tau_k = \Gamma\tau_q = 1.5$ (solid blue line), $\Gamma\tau_k = 0.4, \Gamma\tau_q = 1$ (red dashed line), and $\Gamma\tau_k = 0.2, \Gamma\tau_q = 0.5$ (green dotted line). In panel (b), for $\Gamma_g = \Gamma$, and $\Gamma\tau_k = 0.4, \Gamma\tau_q = 1$, the cases are $\Gamma_{\xi} = 0.3\Gamma$ (solid blue line), $\Gamma_{\xi} = \Gamma$ (red dashed line), and $\Gamma_{\xi} = 3\Gamma$ (green dotted line). The other parameter chosen is $\phi_k = \phi_q = 2n\pi$.}
		\label{fig.3}
\end{figure}
Once $\phi_k \neq 2n\pi$ or $\phi_q \neq 2n\pi$, it can be shown that the atom does not retain a significant excitation in the long-time limit. For the case of $\phi_k = \pi/2, \pi$ and $\phi_q = 2n\pi$, for $t \geq \tau_{\mu}$, the atomic transition $\vert f \rangle \leftrightarrow \vert e \rangle$ is nearly closed, and only photons with wave vector $k$ are emitted to the waveguide through the atomic transition $\vert g \rangle \leftrightarrow \vert e \rangle$. The details of the output field dynamics of the $\Lambda$-type atom are discussed in Section \ref{V}.

~~~~~The case of non-degenerate lower states (corresponding to $\tau_k \neq \tau_q$) is shown in Fig. \ref{fig.3}. We plot the time evolution of the atomic excitation probability $|c_e(t)|^2$ for different values of $\tau_{\mu}$ in panel (a), and for different values of $\Gamma_{\xi}$ with $\Gamma\tau_k = 0.4, \Gamma\tau_q = 0.1$ in panel (b). In Fig. \ref{fig.3}(a), for $\Gamma\tau_k = \Gamma\tau_q = 1.5$ and before $t \geq \tau_{\mu}$, the spontaneous emission of the atom behaves like the purely exponential decay occurring in the case of an infinite waveguide (i.e., the no-mirror case), where the collective decay rate is derived from the sum of the two coupling channels. However, significant changes in the decay rate are observed when $\Gamma\tau_q > \Gamma\tau_k$. Before $t \geq \tau_k$, the atom produces spontaneous emission through both atomic transitions; when $\tau_q \geq t \geq \tau_k$, the decay rate decreases significantly, and the atomic decay is mainly determined by the coupling strength $\Gamma_{\xi}$.
This effect is clearly seen in Fig. \ref{fig.3}(b), where the atomic decay rate satisfies the condition $\dot{c}_e(t)[\Gamma_{\xi} = 3\Gamma] \geq \dot{c}_e(t)[\Gamma_{\xi} = \Gamma] \geq \dot{c}_e(t)[\Gamma_{\xi} = 0.3\Gamma]$ when $\tau_q \geq t \geq \tau_k$.

\section{Discussion of Bound-State Conditions}\label{IV}
	
~~~~~Since the non-Markovian dynamics of an open system are closely related to the energy spectrum of both the system and its environment, investigating the energy spectrum can provide valuable insight into understanding the system's dynamics. Our previous research indicates that when the phase satisfies the condition $\phi_{\mu} = 2n\pi$, a steady state is established during the time evolution. This steady state can be interpreted as the formation of bound states \cite{PhysRevLett.64.2418,tong2010decoherence} between the atom and the photonic environment. The atom-photon bound state can be written as:
$\vert\psi^{bs}\rangle=\int dk c_k^{bs}     
a_k^{\dagger}\vert g,0\rangle+\int dq c_q^{bs}    
a_q^{\dagger}\vert f,0\rangle+c_e^{bs}\vert e,0\rangle$, which must fulfill the normalization condition $\langle \psi^{bs} \vert \psi^{bs} \rangle=1 $ and the time-independent Schrödinger equation $H\vert\psi^{bs}\rangle=E^{bs}\vert\psi^{bs}\rangle$ with $E^{bs}$ denoting bound-state energy, where
	
\begin{equation}\label{eq12}
		E^{bs} c_e^{bs}=\omega_0 c_e^{bs}+\int dk~ gc_k^{bs}+\int dq~ \xi c_q^{bs},
\end{equation}
	
\begin{equation}\label{eq13}
		(E^{bs}-\omega_k) c_k^{bs}=g^{*} c_e^{bs},
\end{equation}

\begin{equation}\label{eq14}
		(E^{bs}-\omega_q-\omega_f) c_q^{bs}=\xi^{*} c_e^{bs}.
\end{equation}
Solving Eqs. (\ref{eq13}) and (\ref{eq14}) leads to
	
\begin{equation}\label{eq15}
		c_k^{bs}=\dfrac{g^{*} c_e^{bs}}{E^{bs}-\omega_0-(k-k_0)v_k},
\end{equation}
	
\begin{equation}\label{eq16}
		c_q^{bs}=\dfrac{\xi^{*} c_e^{bs}}{E^{bs}-\omega_0-(q-q_0)v_q}.
\end{equation}
Together with the normalized condition, we obtain

\begin{eqnarray}\label{eq17}
		|c_e^{bs}|^2&=&\{1+\dfrac{\Gamma_g \tau_k}{2}\mathrm{cos}[(E^{bs}-\omega_0+k_0 v_k )\tau_k]\nonumber \\&& +\dfrac{\Gamma_{\xi} \tau_q}{2}\mathrm{cos}[(E^{bs}-\omega_0+q_0 v_q )\tau_q]\}^{-1}.
\end{eqnarray}
Substituting Eqs. (\ref{eq15}) and (\ref{eq16}) into Eq. (\ref{eq12}), we obtain the transcendental equation for the bound state energy as follows:
	
\begin{eqnarray}\label{eq18}
		E^{bs}&=&\omega_0-\dfrac{\Gamma_g }{2 v_k}\mathrm{sin}[(E^{bs}-\omega_0+k_0 v_k )\tau_k]\nonumber \\&&-\dfrac{\Gamma_{\xi} }{2 v_q} \mathrm{sin}[(E^{bs}-\omega_0+q_0 v_q )\tau_q].
\end{eqnarray}
We can find values of $k$ and $q$ that make the left-hand side of Eqs. (\ref{eq13}) and (\ref{eq14}) vanish. Equivalently, we have $k = (E^{bs} - \omega_0 + k_0 v_k) / v_k$ and $q = (E^{bs} - \omega_0 + q_0 v_q) / v_q$, which leads to $g^{*} = 0$ and $\xi^{*} = 0$, or
\begin{equation}\label{eq19}
		(E^{bs}-\omega_0+k_0 v_k )\tau_k =2n\pi,
\end{equation}
	
\begin{equation}\label{eq20}
		(E^{bs}-\omega_0+q_0 v_q )\tau_q=2n\pi.
\end{equation}
To satisfy the condition of $\phi_{\mu}=\mu_0 \tau_{\mu} v_{\mu}=2n\pi$, we obtain $E^{bs}=\omega_0$, so that $\mu=\mu_0$.
Considering $\vert \psi^{bs}\rangle\langle\psi^{bs}\vert +\int \vert\psi_c (E_c)\rangle\langle\vert\psi_c (E_c)\vert dE_c=I$, we obtain $\vert \psi (t)\rangle=\vert \psi^{bs} (t)\rangle+\vert \psi_c (t)\rangle$, where $\vert \psi^{bs} (t)\rangle$ originates from the contributions of the bound
state with energy $E^{bs}$ given by Eq.(\ref{eq18}), which can be written as
	
\begin{eqnarray}\label{eq21}
		\vert \psi^{bs} (t)\rangle &=&\alpha (E^{bs})e^{-iE^{bs}t}\vert e, 0\rangle+\int dk\beta(E^{bs})e^{-iE^{bs}t}\vert g, 1_k \rangle \nonumber \\&&
		+\int dq\gamma(E^{bs})e^{-iE^{bs}t}\vert f, 1_q \rangle,
\end{eqnarray}
where $\alpha (E^{bs})=|c_e^{bs}|^2$, $\beta(E^{bs})=c_k^{bs}(c_e^{bs})^{*}$ and $\gamma(E^{bs})=c_q^{bs}(c_e^{bs})^{*}$. We show that
$\vert\psi_c(t)\rangle=\int e^{-iE_ct}c_e^c\vert\psi_c(E_c)\rangle dE_c$ is a superposition of the continuous-spectrum eigenfunctions of the Hamiltonian (\ref{eq1})
with the continuous-spectrum eigenfunctions $\vert\psi_c(E_c)\rangle=c_e^c(E_c)\vert e,0\rangle+\int dk c_k^{bs}(E_c)\vert g,1_k\rangle+\int dq c_q^{bs}(E_c)\vert f,1_q\rangle$ \cite{kofman1994spontaneous}, where $E_c$ denotes the continuous-spectrum eigenenergy obtained
by the diagonalization of Hamiltonian (\ref{eq1}). Therefore, the probability amplitude in $\vert e\rangle$ from $\vert \psi (t)\rangle$ is given by
	
\begin{equation}\label{eq22}
		\alpha(E^{bs}) e^{-iE^{bs}t}+\int e^{-iE_ct}\alpha(E_c)dE_c,
\end{equation}
whose integral parts tend to zero in the long-time limit $t\rightarrow\infty$, as stated by the Lebesgue-Riemann lemma \cite{bochner1949fourier}. Therefore,  Eq.(\ref{eq22}) can be simplified to:
	
\begin{equation}\label{eq23}
\alpha(E^{bs})=(1+\dfrac{\Gamma_g\tau_k+\Gamma_{\xi}\tau_q}{2})^{-1}.
\end{equation}
This is precisely consistent with the result obtained from Eq. (\ref{eq11}).

\section{Output Field Dynamics of  the $\Lambda$-type Atom}\label{V}

~~~~~In this section, we focus on the dynamic behavior of the output field of the system in order to measure the light emitted through free space. This provides a natural way to experimentally test the resulting dynamics of the output light at the end of the waveguide. The real-space field annihilation operator at position $x > 0$ can be expressed as:
	
\begin{equation}\label{eq24}
		C(x)=\sum_{\mu=k,q}\sqrt{\dfrac{2}{\pi}}\int d{\mu}~a_{\mu} \mathrm{sin}{\mu} x.
\end{equation}
The prefactor arises from the normalization condition $\int_0^{\infty}dx~C^{\dagger}(x)C(x)=\sum_{\mu=k,q}\int_0^{\infty} d{\mu} a_{\mu}^{\dagger} a_{\mu}$. When applied to the state in Eq.(\ref{eq1}), this gives  $C(x)\vert\Psi(t)\rangle=\psi_k(x, t)\vert g,0\rangle+\psi_q(x, t)\vert f,0\rangle$, where
	
\begin{equation}\label{eq25}
		\psi_\mu (x,t)=\sqrt{\dfrac{2}{\pi}}\int d\mu~c_\mu(\mu,t)\mathrm{sin} \mu x.
\end{equation}
It can be interpreted as the real-space field amplitude. The square modulus of $\psi_\mu(x,t)$ can be measured through the local photon density, which is $\propto \langle \Psi(t) \vert C^{\dagger}(x) C(x) \vert \Psi(t) \rangle = |\psi_k(x,t)|^2 + |\psi_q(x,t)|^2$. We assume that a photon detector is placed at position $\bar{x} = x_0 + d$, where $d > 0$ is the distance between the atom and the detector. One obtains \cite{PhysRevA.87.013820}:
	
\begin{equation}\label{eq26}
		\psi_k(\bar{x},t)=\sqrt{\dfrac{\Gamma_g}{2v_k}}e^{ik_0d}[c_e(t^{'})-e^{i\phi_k}c_e(t^{'}-\tau_k)\Theta(t^{'}-\tau_k)],
\end{equation}

\begin{equation}\label{eq27}
		\psi_q(\bar{x},t)=\sqrt{\dfrac{\Gamma_{\xi}}{2v_q}}e^{iq_0d}[c_e(t^{''})-e^{i\phi_q}c_e(t^{''}-\tau_q)\Theta(t^{''}-\tau_q)],
\end{equation}
where $t^{'}=t-d/v_k$ and $t^{''}=t-d/v_q$.   In the following discussion, we use the symbol $\bar{t}$ to represent both  $t^{'}$ and  $t^{''}$.
	
\begin{figure}[htbp]
		\centering
		\resizebox{0.4\columnwidth}{!}{
			\includegraphics{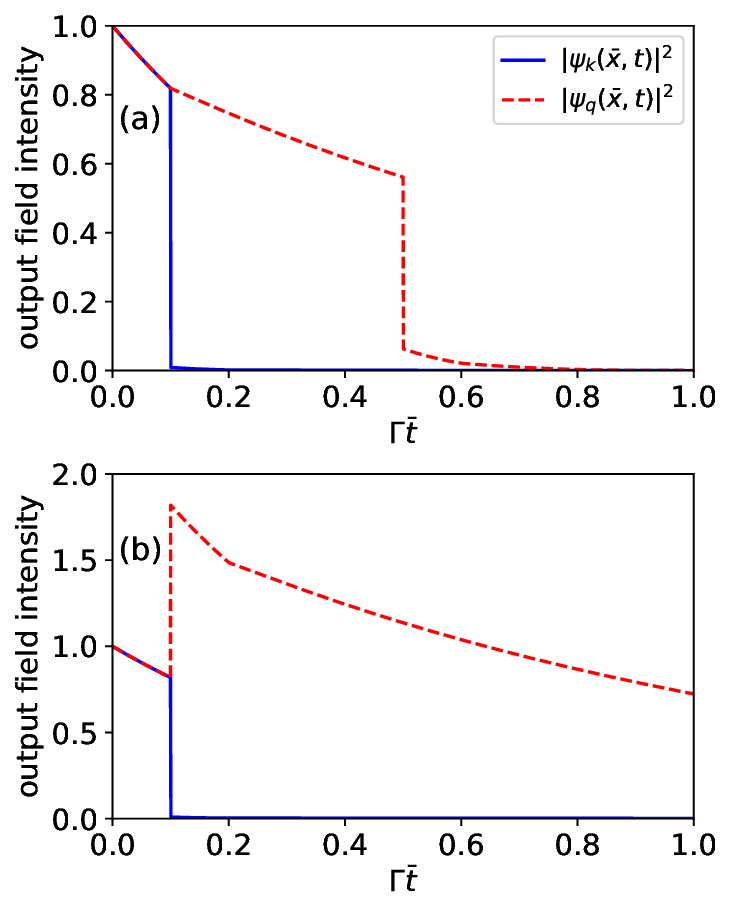}
		}
		\renewcommand\figurename{\textbf{FIG.}}
		\caption[4]{The output field intensities $|\psi_k(\bar{x},t)|^2$ (solid blue line) and $|\psi_q(\bar{x},t)|^2$ (red dashed line) as functions of time $\bar{t}$ (in units of $\Gamma^{-1}$) are shown for the following cases: (a) $\phi_k = \phi_q = 2n\pi$, $\Gamma \tau_k = 0.1$, $\Gamma \tau_q = 0.5$; (b) $\phi_k = 2n\pi$, $\phi_q = \pi/2$, $\Gamma \tau_k = \Gamma \tau_q = 0.1$. The other parameters is chosen as $\Gamma_g = \Gamma_{\xi} = \Gamma$.}
		\label{fig.4}
\end{figure}

\begin{figure}[htbp]
		\centering
		\resizebox{0.4\columnwidth}{!}{
			\includegraphics{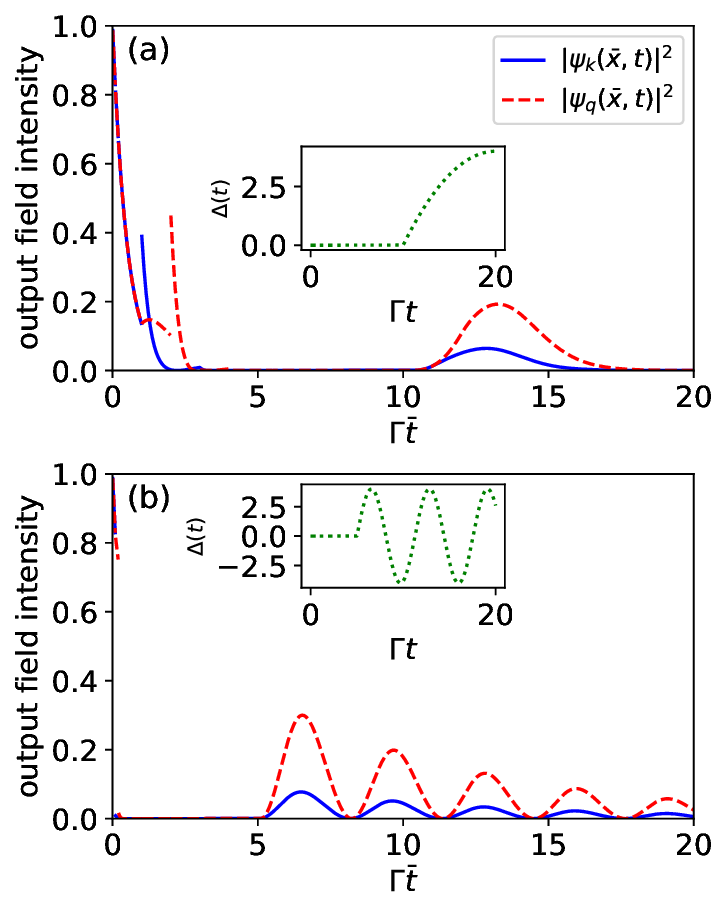}
		}
		\renewcommand\figurename{\textbf{FIG.}}
		\caption[5]{The output field intensities $|\psi_k(\bar{x},t)|^2$ (solid blue line) and $|\psi_q(\bar{x},t)|^2$ (red dashed line) as functions of time $\bar{t}$ (in units of $\Gamma^{-1}$) are shown under an applied frequency shift $\Delta(t)$ (green dotted line) for the following cases: (a) $\Gamma \tau_k = 1$, $\Gamma \tau_q = 2$; (b) $\Gamma \tau_k = 0.1$, $\Gamma \tau_q = 0.2$. The other parameters are $\Gamma_g = \Gamma_{\xi} = \Gamma$ and $\phi_k = \phi_q = 2n\pi$.}
		\label{fig.5}
\end{figure}

~~~~~In Fig. \ref{fig.4}, we plot the output field intensity $|\psi_{\mu}(\bar{x},t)|^2$ as a function of time $\bar{t}$. In Fig. \ref{fig.4} (a), for the case where $\phi_k = \phi_q = 2n\pi$, and for $t \geq \tau_k$, both atomic transitions $\vert g \rangle \leftrightarrow \vert e \rangle$ and $\vert f \rangle \leftrightarrow \vert e \rangle$ are coupled to the waveguide, emitting photons with wave vectors $k$ and $q$ to the output port synchronously. When $t$ is between $\tau_k$ and $\tau_q$, $|\psi_k(\bar{x},t)|^2$ rapidly decays to nearly zero, while $|\psi_q(\bar{x},t)|^2$ remains large. This phenomenon indicates that, during this period, the excited $\Lambda$-type atom undergoes spontaneous emission through the atomic transition $\vert f \rangle \leftrightarrow \vert e \rangle$. However, for $t \geq \tau_q$, the presence of the mirror starts to influence the atom, causing the atom to retain a significant amount of excitation even as $t \to \infty$, as discussed in Section \ref{III}. In Fig. \ref{fig.4} (b), where $\phi_k = 2n\pi$ and $\phi_q = \pi/2$, it is clear that the instantaneous and retarded decay rates of $\Gamma_{\xi}$ do not reach equilibrium. Therefore, for $t \geq \tau_{\mu}$, photons with wave vector $q$ are continuously emitted into the waveguide until the excited atom decays completely to the ground state.

\begin{figure}[htbp]
		\centering
		\resizebox{0.4\columnwidth}{!}{
			\includegraphics{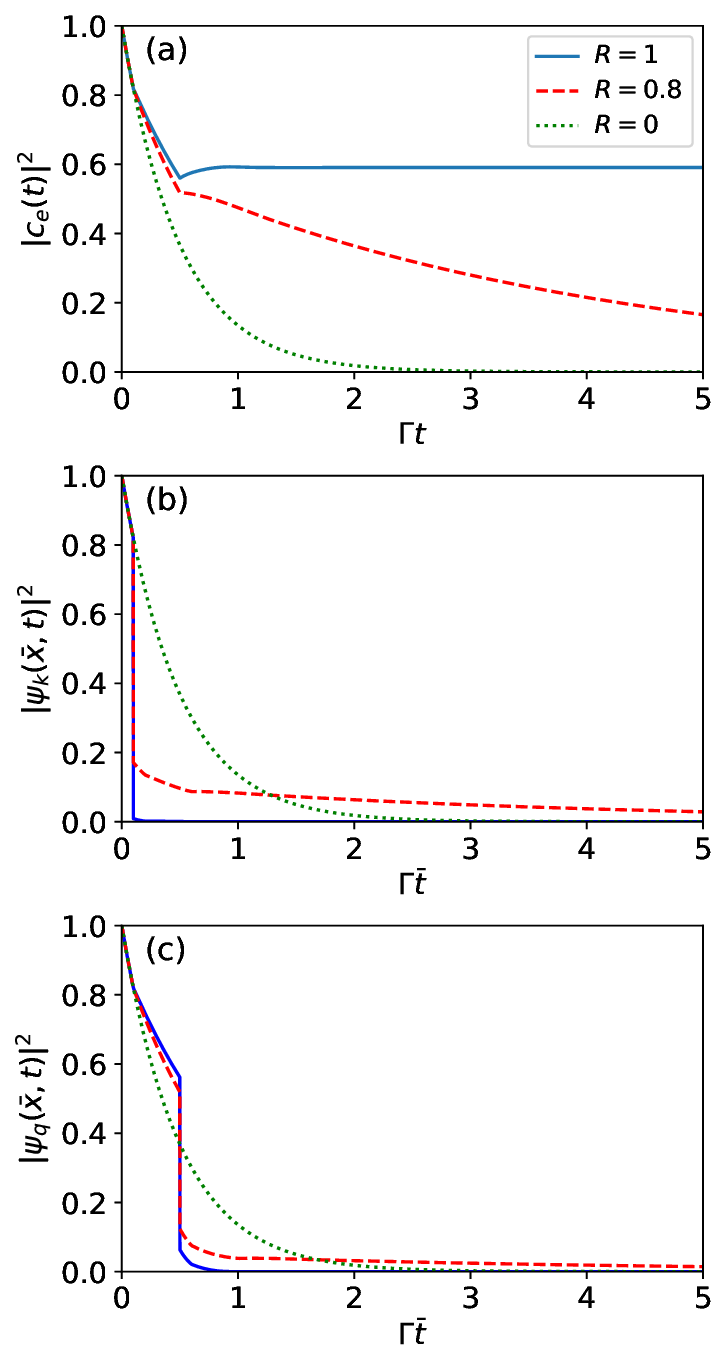}
		}
		\renewcommand\figurename{\textbf{FIG.}}
		\caption[6]{Robustness of $|c_e(t)|^2$ in (a) as a function of time $t$ (in units of $\Gamma^{-1}$). Robustness of $|\psi_k(\bar{x},t)|^2$ in (b) and $|\psi_q(\bar{x},t)|^2$ in (c) as functions of time $\bar{t}$ (in units of $\Gamma^{-1}$). The cases of $R = 1$ (solid blue line), $R = 0.8$ (red dashed line), and $R = 0$ (green dotted line) are shown. The other parameters are $\Gamma_g = \Gamma_{\xi} = \Gamma$, $\phi_k = \phi_q = 2n\pi$, $\Gamma \tau_k = 0.1$, and $\Gamma \tau_q = 0.5$.}
		\label{fig.6}
\end{figure}
	
~~~~~As discussed previously, when the phase $\phi_{\mu} = 2n\pi$, the spontaneous emission of the atom is suppressed in the long-time limit. However, when a frequency shift is applied to the atom, it forces the trapped excitation to be released. In Fig. \ref{fig.5}, we model the frequency shift as a smooth time function $\Delta(t)$ (i.e., $\dot{c}_e(t) \rightarrow \dot{c}_e(t) + i\Delta(t) c_e(t)$) and plot the resulting numerically computed output field intensity as a function of time. Two types of wave functions for $\Delta(t)$ are shown in the insets. As illustrated in Fig. \ref{fig.5} (a), before the atomic frequency shift $\Delta(t)$ is applied, the output field intensities $|\psi_k(\bar{x},t)|^2$ and $|\psi_q(\bar{x},t)|^2$ decay to zero sequentially, at times $t = \tau_k$ and $t = \tau_q$, respectively. Once $\Delta(t) \neq 0$, photons with wave vectors $k$ and $q$ are re-emitted into the waveguide with different output field intensities. The difference in intensity arises from the retarded decay rates $\Gamma_g$ and $\Gamma_{\xi}$. When $\Delta(t)$ oscillates at a high frequency, as shown in Fig. \ref{fig.5} (b), the output field intensity exhibits a continuous oscillation with decreasing amplitude until it eventually decays to zero.

\section{Discussion  of the Reflectivity $R<1$ Condition}\label{VI}
	
~~~~~So far, we have discussed the case where the mirror is perfect (i.e., $R = 1$). To assess the experimental observability, we now assume that the waveguide is terminated at $x = 0$ by a non-ideal mirror with reflectivity $R < 1$. The presence of an imperfect mirror modifies the delay term in Eq. (\ref{eq6}), as the atom will re-interact only with the portion of light that is reflected. This leads to the substitution $e^{i\phi_{\mu}} \rightarrow r e^{i\phi_{\mu}}$, where $r$ is the complex probability amplitude for backward reflection off the mirror ($r = R + i\sqrt{R(1 - R)}$). Finally, as mentioned in the main text, the modified equations for the probability amplitude and output field intensity are expressed as:
	
\begin{equation}\label{eq28}
		\begin{aligned}
			&\dot{c}_{e}(t)=-\frac{\Gamma_g}{2}c_{e}(t)+r\frac{\Gamma_g}{2}e^{i\phi_k}c_{e}(t-\tau_k)\Theta(t-\tau_k)\\
			&~~~~~~~~~-\frac{\Gamma_{\xi}}{2}c_{e}(t)+r\frac{\Gamma_{\xi}}{2}e^{i\phi_q}c_{e}(t-\tau_q)\Theta(t-\tau_q),
		\end{aligned}
\end{equation}

\begin{equation}\label{eq29}
		\psi_k(\bar{x},t)=\sqrt{\dfrac{\Gamma_g}{2v_k}}e^{ik_0d}[c_e(t^{'})-re^{i\phi_k}c_e(t^{'}-\tau_k)\Theta(t^{'}-\tau_k)],
\end{equation}

\begin{equation}\label{eq30}
		\psi_q(\bar{x},t)=\sqrt{\dfrac{\Gamma_{\xi}}{2v_q}}e^{iq_0d}[c_e(t^{''})-re^{i\phi_q}c_e(t^{''}-\tau_q)\Theta(t^{''}-\tau_q)].
\end{equation}
	
~~~~~We simulated the ideal case ($R = 1$), the no-mirror case ($R = 0$), and the intermediate case ($R = 0.8$) in Fig. \ref{fig.6}. For a non-ideal semi-infinite waveguide, the long-lived population is not achieved in the long-time limit, as shown in Fig. \ref{fig.6} (a), and the decay rate increases as $R$ decreases after the time delay. This dynamic process can be further observed from the output field intensity in Figs. \ref{fig.6} (b) and (c). For $t \geq \tau_{\mu}$, the intensity of $|\psi_{\mu}(\bar{x}, t)|^2$ is significantly reduced, but not to zero. The atom continues to emit photons into the waveguide, since the retarded decay rates are smaller than the instantaneous decay rates $\Gamma_{\mu}$, and the two parts will not reach equilibrium until the atom fully decays to the ground state.

\section{Two $\Lambda$-type Atom Coupled to The Single-end Photonic Waveguide}\label{VII}
	
~~~~~We extend the model to the case of two $\Lambda$-type atoms in a semi-infinite waveguide system, as shown in Fig. \ref{fig.1} (b), to explore the disentanglement dynamics of the two separate atoms. This configuration is equivalent to the topological form of nested giant atoms in an infinite waveguide \cite{PhysRevLett.120.140404}. The corresponding Hamiltonian is expressed as:
	
\begin{eqnarray}\label{eq31} 
		H &=&\sum_{j}^{2}(\omega_0\vert e_j\rangle\langle e_j\vert +\omega_f\vert f_j\rangle\langle f_j\vert) +\sum_{\mu=k,q}\int_{0}^{\mu_c} d\mu ~\omega_\mu a_\mu^{\dagger}a_\mu\nonumber \\&&+\sum_{j}^{2}(\int_{0}^{\kappa_c} dk ~g\vert e_j\rangle\langle g_j\vert a_k+\int_{0}^{q_c} dq ~\xi\vert e_j\rangle\langle f_j\vert a_q+H.c.).
\end{eqnarray}
The state of the total system in the single excitation subspace can be written as
\begin{eqnarray}\label{eq32} 
		\vert \psi(t) \rangle &=&\int dk~ c_{k}(t)     
		a_k^{\dagger}\vert g,0\rangle+\int dq~ c_{q}(t)     
		a_q^{\dagger}\vert f,0\rangle\nonumber \\&&+\sum_{j}^{2}c_{ei}(t)\vert e_j,0\rangle.
\end{eqnarray}
By solving the time-dependent Schrödinger equation, there are
	
\begin{equation}\label{eq33}
		\begin{aligned}
			&\dot{c}_{e1}(t)=-i\omega_0 c_{e1}(t)-i\int_{0}^{\kappa_c} dk~ gc_{k}(t)-i\int_{0}^{q_c} dq~ \xi c_{q}(t),\\
			&\dot{c}_{e2}(t)=-i\omega_0 c_{e2}(t)-i\int_{0}^{\kappa_c} dk~ gc_{k}(t)-i\int_{0}^{q_c} dq~ \xi c_{q}(t),\\
			&\dot{c}_{k}(t)=-i\omega_k c_{k}(t)-\sum_{j}^{2}ig^{*} c_{ej}(t),\\
			&\dot{c}_{q}(t)=-i(\omega_q+\omega_f) c_{q}(t)-\sum_{j}^{2}i\xi^{*} c_{ej}(t),\\
		\end{aligned}
\end{equation}
for the case of $c_{k}(0)=c_{q}(0)=0$, the formal solution of $c_k(t)$ and $c_q(t)$ can be written as
	
\begin{equation}\label{eq34}
		\begin{aligned}
			&c_{k}(t)=-i\int_0^{t} dt^{'}e^{-iv_k(k-k_0)(t-t^{'})}\\
			&~~~~~~~~~\times\sqrt{\Gamma_g v_k/\pi} [\mathrm{sin}(\kappa x_1)c_{e1}(t^{'})+\mathrm{sin}(\kappa x_2)c_{e2}(t^{'})],\\
			&c_{q}(t)=-i\int_0^{t} dt^{'}e^{-iv_q(q-q_0)(t-t^{'})}\\
			&~~~~~~~~~\times\sqrt{\Gamma_{\xi} v_q/\pi} [\mathrm{sin}(q x_1)c_{e1}(t^{'})+\mathrm{sin}(q x_2)c_{e2}(t^{'})].
		\end{aligned}
\end{equation}
	
~~~~~Substituting Eq.(\ref{eq34}) into the dynamic equation of $c_e(t)$ in Eq.(\ref{eq33}), there is
	
\begin{equation}\label{eq35}
		\begin{aligned}
			&\dot{c}_{e1}(t)=-\frac{\Gamma_g v_k}{\pi}\int_0^t dt^{'}c_{e1}(t^{'})e^{ik_0v_k(t-t^{'})}\int_{-\infty}^{+\infty} dk~ \mathrm{sin}^2(k x_1)\\
			&~~~~~~~~~~~\times e^{-ikv_k(t-t^{'})}\\
			&~~~~~~~~~~~-\frac{\Gamma_g v_k}{\pi}\int_0^t dt^{'}c_{e2}(t^{'})e^{ik_0v_k(t-t^{'})}\\ 
			&~~~~~~~~~~~\times \int_{-\infty}^{+\infty} dk~ \mathrm{sin}(k x_1)\mathrm{sin}(k x_2) e^{-ikv_k(t-t^{'})}\\
			&~~~~~~~~~~~-\frac{\Gamma_{\xi} v_q}{\pi}\int_0^t dt^{'}c_{e1}(t^{'})e^{iq_0v_q(t-t^{'})}\int_{-\infty}^{+\infty} dq~ \mathrm{sin}^2(q x_1)\\
			&~~~~~~~~~~~\times e^{-iqv_q(t-t^{'})}\\
			&~~~~~~~~~~~-\frac{\Gamma_{\xi} v_q}{\pi}\int_0^t dt^{'}c_{e2}(t^{'})e^{iq_0v_q(t-t^{'})}\\ 
			&~~~~~~~~~~~\times \int_{-\infty}^{+\infty} dq~ \mathrm{sin}(q x_1)\mathrm{sin}(q x_2) e^{-iqv_q(t-t^{'})}.
		\end{aligned}
\end{equation}
After integrating over the wave vector $\mu$ and time $t'$ in Eq. (\ref{eq35}), one can obtain a delayed differential equation for $c_{ej}(t)$ (see \hyperref[Appendix]{Appendix} for more details):
\begin{equation}\label{eq36}
		\begin{aligned}
			&\dot{c}_{e1}(t)=-\frac{\Gamma_g}{2}c_{e1}(t)+\frac{\Gamma_g}{2}e^{i\phi_{k1}}c_{e1}(t-\tau_{k1})\Theta(t-\tau_{k1})\\
			&~~~~~~~~~~-\frac{\Gamma_g}{2}e^{i\phi_{k-}}c_{e2}(t-\tau_{k-})\Theta(t-\tau_{k-})\\
			&~~~~~~~~~~+\frac{\Gamma_g}{2}e^{i\phi_{k+}}c_{e2}(t-\tau_{k+})\Theta(t-\tau_{k+})\\
			&~~~~~~~~~~-\frac{\Gamma_{\xi}}{2}c_{e1}(t)+\frac{\Gamma_{\xi}}{2}e^{i\phi_{q1}}c_{e1}(t-\tau_{q1})\Theta(t-\tau_{q1})\\
			&~~~~~~~~~~-\frac{\Gamma_{\xi}}{2}e^{i\phi_{q-}}c_{e2}(t-\tau_{q-})\Theta(t-\tau_{q-})\\
			&~~~~~~~~~~+\frac{\Gamma_{\xi}}{2}e^{i\phi_{q+}}c_{e2}(t-\tau_{q+})\Theta(t-\tau_{q+}),
		\end{aligned}
\end{equation}
and
\begin{equation}\label{eq37}
		\begin{aligned}
			&\dot{c}_{e2}(t)=-\frac{\Gamma_g}{2}c_{e2}(t)+\frac{\Gamma_g}{2}e^{i\phi_{k2}}c_{e2}(t-\tau_{k2})\Theta(t-\tau_{k2})\\
			&~~~~~~~~~~-\frac{\Gamma_g}{2}e^{i\phi_{k-}}c_{e1}(t-\tau_{k-})\Theta(t-\tau_{k-})\\
			&~~~~~~~~~~+\frac{\Gamma_g}{2}e^{i\phi_{k+}}c_{e1}(t-\tau_{k+})\Theta(t-\tau_{k+})\\
			&~~~~~~~~~~-\frac{\Gamma_{\xi}}{2}c_{e2}(t)+\frac{\Gamma_{\xi}}{2}e^{i\phi_{q2}}c_{e2}(t-\tau_{q2})\Theta(t-\tau_{q2})\\
			&~~~~~~~~~~-\frac{\Gamma_{\xi}}{2}e^{i\phi_{q-}}c_{e1}(t-\tau_{q-})\Theta(t-\tau_{q-})\\
			&~~~~~~~~~~+\frac{\Gamma_{\xi}}{2}e^{i\phi_{q+}}c_{e1}(t-\tau_{q+})\Theta(t-\tau_{q+}),
		\end{aligned}
\end{equation}
where $\tau_{\mu j}=2x_j/v_{\mu}$, $\tau_{\mu -}=(x_2-x_1)/v_{\mu}$, $\tau_{\mu +}=(x_2+x_1)/v_{\mu}$ and  $\phi_{\mu j}=2x_j \mu_0$, $\phi_{\mu -}=(x_2-x_1)\mu_0$, $\phi_{\mu +}=(x_2+x_1)\mu_0$.
	
~~~~~Now, we consider the specific case of two atoms at different positions. When $x_1 \rightarrow 0$, the first atom is placed near the mirror at the end of the waveguide. In this case, we have $\tau_{k1} \approx \tau_{q1} \approx 0$, $\tau_{k-} \approx \tau_{k+}$, $\tau_{q-} \approx \tau_{q+}$, and $\phi_{k1} \approx \phi_{q1} \approx 0$, $\phi_{k-} \approx \phi_{k+}$, $\phi_{q-} \approx \phi_{q+}$. Therefore, Eqs. (\ref{eq36}) and (\ref{eq37}) simplify to:
\begin{equation}\label{eq38}
		\dot{c}_{e1}(t)=0,
\end{equation}
\begin{equation}\label{eq39}
	\begin{aligned}
		&\dot{c}_{e2}(t)=-\frac{\Gamma_g}{2}c_{e2}(t)+\frac{\Gamma_g}{2}e^{i\phi_{k2}}c_{e2}(t-\tau_{k2})\Theta(t-\tau_{k2})\\
		&~~~~~~~~~~-\frac{\Gamma_{\xi}}{2}c_{e2}(t)+\frac{\Gamma_{\xi}}{2}e^{i\phi_{q2}}c_{e2}(t-\tau_{q2})\Theta(t-\tau_{q2}).
	\end{aligned}
\end{equation}
Under what conditions is the first atom always in a state where spontaneous emission is suppressed, while the dynamics of the second atom resemble the single-atom case discussed in Section \ref{III}. When $x_1 = x_2 = x_0$ (i.e., the positions of the two atoms coincide), there is:
	
\begin{equation}\label{eq40}
	\begin{aligned}
		&\dot{c}_{e1}(t)=\dot{c}_{e2}(t)\\
		&~~~~~~~=-\frac{\Gamma_g+\Gamma_{\xi}}{2}[c_{e1}(t)+c_{e2}(t)]\\
		&~~~~~~~~~~+\frac{\Gamma_g}{2}e^{i\phi_k}[c_{e1}(t-\tau_k)\Theta(t-\tau_k)+c_{e2}(t-\tau_k)\Theta(t-\tau_k)]\\
		&~~~~~~~~~~+\frac{\Gamma_{\xi}}{2}e^{i\phi_q}[c_{e1}(t-\tau_q)\Theta(t-\tau_q)+c_{e2}(t-\tau_q)\Theta(t-\tau_q)],
	\end{aligned}
\end{equation}
In the single-excitation case, the two atoms can be assumed to be in a general entangled state, $\vert \Psi \rangle_{\pm} = (\vert e \rangle_1 \vert g \rangle_2 \pm \vert g \rangle_1 \vert e \rangle_2) / \sqrt{2}$, which corresponds to the symmetric and antisymmetric states: $p_+(t) = [c_{e1}(t) + c_{e2}(t)] / \sqrt{2}$ and $p_-(t) = [c_{e1}(t) - c_{e2}(t)] / \sqrt{2}$, respectively. The equations of motion for the amplitudes of the symmetric and antisymmetric states can then be derived as:
	
\begin{equation}\label{eq41}
		\begin{aligned}
			&\dot{P}_{+}(t)=[-\Gamma_g P_{+}(t)+\Gamma_g e^{i\phi_k}P_{+}(t-\tau_k)\Theta (t-\tau_k)\\
			&~~~~~~~~~~-\Gamma_{\xi} P_{+}(t)+\Gamma_{\xi} e^{i\phi_q}P_{+}(t-\tau_q)\Theta (t-\tau_q)]/\sqrt{2},\\
			&\dot{P}_{-}(t)=0.
		\end{aligned}
\end{equation}
Obviously, the equation of motion for the symmetric state is similar to Eqs. (\ref{eq6}) and (\ref{eq39}). In contrast, for the antisymmetric state, the atoms are in a dark state, as the disentanglement dynamics of the collective system is independent of the phases.
	
\section{Conclusion}\label{VIII}
	
~~~~~We have investigated the non-Markovian dynamics of spontaneous emission for a $\Lambda$-type atom coupled to a semi-infinite 1D photonic waveguide through two atomic transitions, $\vert g \rangle \leftrightarrow \vert e \rangle$ and $\vert f \rangle \leftrightarrow \vert e \rangle$, with photons emitted at wave vectors $k$ and $q$, respectively. Based on the exact delayed differential equation for the atomic excitation amplitude discussed in the main text, the decay rate of the excited atom is entirely determined by the two coupling channels. The spontaneous emission process is suppressed only when the emitted photons of the two wavelengths match the atom-mirror distance (i.e., $\phi_{\mu} = 2n\pi$), ensuring the formation of atom-photon bound states.
Moreover, the atomic system will emit specific photons in a single mode when one of the coupling channels is blocked. We have explored the output field intensity to verify the entailed dynamics of such output light, even when $\phi_{\mu} = 2n\pi$, by applying a frequency shift to the atom. The mirror in this system acts as a feedback mechanism, and when the mirror is not ideal (i.e., $R < 1$), simulations show that the instantaneous and retarded decay rates of $\Gamma_{g}$ and $\Gamma_{\xi}$ will not reach equilibrium.
	
~~~~~We also explore the disentanglement dynamics of the two separate atoms and demonstrate that: (1) when the first atom is placed near the mirror, it suppresses spontaneous emission, and the dynamics of the second atom resemble those of the single-atom case; (2) when the positions of the two atoms coincide, we divide the collective atomic system into symmetric and antisymmetric states. The decay of the symmetric state depends on the phase, while the antisymmetric state is independent of the phase and remains in a dark state.

\appendix
\renewcommand{\theequation}{\arabic{equation}}
\setcounter{equation}{0}
	
\section*{Appendix: Detailed Solution of Eqs. (\ref{eq6}) and (\ref{eq36})}\label{Appendix}
	
~~~~~Consider converting the integral over the photon wave vector to an integral over the frequency. The new differential form of Eq. (\ref{eq5}) is then represented as:
	
\begin{equation}\label{eqa1}
		\begin{aligned}
			&\dot{c}_{e}(t)=-\frac{\Gamma_g}{4\pi}\int_0^t dt^{'}c_{e}(t^{'})e^{ik_0 v_k(t-t^{'})}\int_{-\infty}^{+\infty} d(kv_k)\\
			&~~~~~~~~\times[2e^{-ikv_k(t-t^{'})}-e^{-ikv_k(t-t^{'}+\tau_k)}-e^{-ikv_ k(t-t^{'}-\tau_k)}]\\
			&~~~~~~~~-\frac{\Gamma_{\xi}}{4\pi}\int_0^t dt^{'}c_{e}(t^{'})e^{iq_0 v_q(t-t^{'})}\int_{-\infty}^{+\infty} d(qv_q)\\
			&~~~~~~~~\times[2e^{-iqv_q(t-t^{'})}-e^{-iqv_q(t-t^{'}+\tau_q)}-e^{-iqv_q(t-t^{'}-\tau_q)}].
		\end{aligned}
\end{equation}
Using the definition $\int d\omega~\mathrm{exp}(i\omega t)=2\pi \delta(t)$ of the $\delta$ function and its sifting property  $\int f(t)\delta(t-t^{'})=f(t^{'})$ Eq. (\ref{eqa1}) can be further simplified to
	
\begin{equation}\label{eqa2}
		\begin{aligned}
			&\dot{c}_{e}(t)=-\frac{\Gamma_g}{2}\int_0^t dt^{'}c_{e}(t^{'})e^{-ik_0 v_k(t-t^{'})}\\
			&~~~~~~~~~\times[2\delta(t-t^{'})-\delta(t-t^{'}+\tau_k)-\delta(t-t^{'}-\tau_k)]\\
			&~~~~~~~~~-\frac{\Gamma_{\xi}}{2}\int_0^t dt^{'}c_{e}(t^{'})e^{-iq_0 v_q(t-t^{'})}\\
			&~~~~~~~~~\times[2\delta(t-t^{'})-\delta(t-t^{'}+\tau_q)-\delta(t-t^{'}-\tau_q)]\\
			&~~~~~=-\frac{\Gamma_g}{2}c_{e}(t)+\frac{\Gamma_g}{2}e^{i\phi_k}c_{e}(t-\tau_k)\Theta(t-\tau_k)\\
			&~~~~~~~~-\frac{\Gamma_{\xi}}{2}c_{e}(t)+\frac{\Gamma_{\xi}}{2}e^{i\phi_q}c_{e}(t-\tau_q)\Theta(t-\tau_q).
		\end{aligned}
\end{equation}
We have discarded the time-advanced term containing $\delta(t-t^{'}+\tau_{\mu})$, as it contributes zero to the integral.
	
Following the same procedure and generalizing to the two-atom case shown in Fig. (\ref{fig.1}) (b) in the main text, we have:

\begin{equation}\label{eqa3}
		\begin{aligned}
			&\dot{c}_{e1}(t)=-\frac{\Gamma_g v_k}{\pi}\int_0^t dt^{'}c_{e1}(t^{'})e^{ik_0v_k(t-t^{'})}\int_{-\infty}^{+\infty} dk~ \mathrm{sin}^2(k x_1) e^{-ikv_k(t-t^{'})}-\frac{\Gamma_g v_k}{\pi}\int_0^t dt^{'}c_{e2}(t^{'})e^{ik_0v_k(t-t^{'})}\\ 
			&~~~~~~~~\times \int_{-\infty}^{+\infty} dk~ \mathrm{sin}(k x_1)\mathrm{sin}(k x_2) e^{-ikv_k(t-t^{'})}\\
			&~~~~~~~~-\frac{\Gamma_{\xi} v_q}{\pi}\int_0^t dt^{'}c_{e1}(t^{'})e^{iq_0v_q(t-t^{'})}\int_{-\infty}^{+\infty} dq~ \mathrm{sin}^2(q x_1)e^{-iqv_q(t-t^{'})}-\frac{\Gamma_{\xi} v_q}{\pi}\int_0^t dt^{'}c_{e2}(t^{'})e^{iq_0v_q(t-t^{'})}\\ 
			&~~~~~~~~\times \int_{-\infty}^{+\infty} dq~ \mathrm{sin}(q x_1)\mathrm{sin}(q x_2) e^{-iqv_q(t-t^{'})}\\
			&~~~~=-\frac{\Gamma_g}{4\pi}\int_0^t dt^{'}c_{e1}(t^{'})e^{ik_0v_k(t-t^{'})}\int_{-\infty}^{+\infty} d(kv_k)[2e^{-ikv_k(t-t^{'})}-e^{-ikv_k(t-t^{'}+\tau_{k1})}-e^{-ikv_k(t-t^{'}-\tau_{k1})}]\\
			&~~~~~-\frac{\Gamma_g}{4\pi}\int_0^t dt^{'}c_{e2}(t^{'})e^{ik_0v_k(t-t^{'})}\int_{-\infty}^{+\infty} d(kv_k)[e^{-ikv_k(t-t^{'}+\tau_{k-})}+e^{-ikv_k(t-t^{'}-\tau_{k-})}-e^{-ikv_k(t-t^{'}+\tau_{k+})}-e^{-ikv_k(t-t^{'}-\tau_{k+})}]\\
			&~~~~~~~~-\frac{\Gamma_{\xi}}{4\pi}\int_0^t dt^{'}c_{e1}(t^{'})e^{iq_0v_q(t-t^{'})}\int_{-\infty}^{+\infty} d(qv_q)[2e^{-iqv_q(t-t^{'})}-e^{-iqv_q(t-t^{'}+\tau_{q1})}-e^{-iqv_q(t-t^{'}-\tau_{q1})}]\\
			&~~~~~~~~-\frac{\Gamma_{\xi}}{4\pi}\int_0^t dt^{'}c_{e2}(t^{'})e^{iq_0v_q(t-t^{'})}\int_{-\infty}^{+\infty} d(qv_q)[e^{-iqv_q(t-t^{'}+\tau_{q-})}+e^{-iqv_q(t-t^{'}-\tau_{q-})}-e^{-iqv_q(t-t^{'}+\tau_{q+})}-e^{-iqv_q(t-t^{'}-\tau_{q+})}]\\
			&~~~~=-\frac{\Gamma_g}{2}\int_0^t dt^{'}c_{e1}(t^{'})e^{ik_0v_k(t-t^{'})}[2\delta(t-t^{'})-\delta(t-t^{'}+\tau_{k1})-\delta(t-t^{'}-\tau_{k1})]\\
			&~~~~~~~~-\frac{\Gamma_g}{2}\int_0^t dt^{'}c_{e2}(t^{'})e^{ik_0v_k(t-t^{'})}[\delta(t-t^{'}+\tau_{k-})+\delta(t-t^{'}-\tau_{k-})-\delta(t-t^{'}+\tau_{k+})-\delta(t-t^{'}-\tau_{k+})]\\
			&~~~~~~~~-\frac{\Gamma_{\xi}}{2}\int_0^t dt^{'}c_{e1}(t^{'})e^{iq_0v_q(t-t^{'})}[2\delta(t-t^{'})-\delta(t-t^{'}+\tau_{q1})-\delta(t-t^{'}-\tau_{q1})]\\
			&~~~~~~~~-\frac{\Gamma_{\xi}}{2}\int_0^t dt^{'}c_{e2}(t^{'})e^{iq_0v_q(t-t^{'})}[\delta(t-t^{'}+\tau_{q-})+\delta(t-t^{'}-\tau_{q-})-\delta(t-t^{'}+\tau_{q+})-\delta(t-t^{'}-\tau_{q+})]\\
			&~~~~=-\frac{\Gamma_g}{2}c_{e1}(t)+\frac{\Gamma_g}{2}e^{i\phi_{k1}}c_{e1}(t-\tau_{k1})\Theta(t-\tau_{k1})-\frac{\Gamma_g}{2}e^{i\phi_{k-}}c_{e2}(t-\tau_{k-})\Theta(t-\tau_{k-})+\frac{\Gamma_g}{2}e^{i\phi_{k+}}c_{e2}(t-\tau_{k+})\Theta(t-\tau_{k+})\\
			&~~~~~~~-\frac{\Gamma_{\xi}}{2}c_{e1}(t)+\frac{\Gamma_{\xi}}{2}e^{i\phi_{q1}}c_{e1}(t-\tau_{q1})\Theta(t-\tau_{q1})-\frac{\Gamma_{\xi}}{2}e^{i\phi_{q-}}c_{e2}(t-\tau_{q-})\Theta(t-\tau_{q-})+\frac{\Gamma_{\xi}}{2}e^{i\phi_{q+}}c_{e2}(t-\tau_{q+})\Theta(t-\tau_{q+}).
		\end{aligned}
\end{equation}

\medskip
\textbf{Acknowledgements} \par 
This work is supported by the National Natural Science
Foundation of China under Grant No. 12375018.

\medskip
\textbf{Conflict of Interest} \par 
The authors declare no conflict of interest.

\medskip
\textbf{Data Availability Statement} \par 
The data that support the findings of this study are available from the corresponding author upon reasonable request.

\medskip
\textbf{Keywords} \par 
Non-Markovian dynamics, single-end photonic waveguide, interference, spontaneous emission

\medskip
	
	%
	

\end{document}